# Initial imprint effect on dynamic mitigation of plasma instability

S. Kawata
Graduate School of Engineering, Utsunomiya University,
Yohtoh 7-1-2, Utsunomiya 321-8585, Japan.

We proposed a dynamic mitigation method for plasma instabilities based on a phase control to mitigate plasma instabilities and to smooth plasma non-uniformities [e.g. Phys. Plasmas, 19(2012)024503]. In plasmas, perturbation phase would be unknown in general, and instability growth rate is discussed. However, if the perturbation is introduced by, for example, an illumination non-uniformity of an input energy driver beam, the perturbation phase would be defined by the driver illumination non-uniformity itself. When the driver axis is controlled by its axis oscillation or wobbling motion, the perturbation phase would be known and controlled. By the superimposition of the growing phase-controlled perturbations, the overall plasma instability growth is mitigated. The dynamic mitigation method is effective to mitigate growths of various plasma instabilities. At the same time, it was found that the phase of the growing perturbations mitigated would be still defined by the initial imprint. In this paper, the initial imprint effect is focused on the dynamic mitigation mechanism in plasmas. The results in this paper demonstrate that the initial imprint effect is reduced by a pulse shaping of the oscillating or wobbling perturbation.

## I. INTRODUCION

Plasma stability has been one of central concerns in plasma physics and applications, including fusion plasmas, space plasmas, discharge plasmas, industrial application plasmas, etc.[1-6] Plasma perturbation would feature the instability onset. Normally perturbation phase is unknown in plasmas, and so instability growth rate is discussed. On the other hand, if the perturbation phase is defined, for example, by an externally input energy driver, the instability growth would be controlled by a superimposition of perturbations introduced actively.[6-8] This mechanism may be similar to feedforward control method, which is widely used to stabilize society systems.[9] In plasmas perturbation phase and amplitude cannot be measured dynamically. However, dynamically controlled drivers may introduce and define the initial perturbations actively. For example, heavy ion beam (HIB) accelerators provide a well-established capability to oscillate a HIB axis with a high frequency.[10-13] An intense HIB axis can be wobbled in its controlled way, and the HIB energy deposition in plasmas defines the perturbation phase and amplitude. In this case, plasma instability growth would be controlled by the superimposition of perturbations imposed actively at least in its linear phase.

For example, an input energy driver delivered to a plasma would introduce a perturbation with its bulk energy deposition. When the driver oscillates or wobbles in a controlled way, the perturbation phase would be also oscillated or wobbled. At each time the driver introduces a new perturbation with its own phase, and the overall instability amplitude becomes the superimposition of all the successive growing perturbations. One example may be a tearing mode instability driven by a sheet current. If the sheet current is oscillated or wobbled, for example, along the sheet, the tearing instability growth would be mitigated.[6, 14] Another example may be the dynamic mitigation of the Rayleigh-Taylor instability (RTI) growth.[6-8, 15] The RTI would prevent a uniform fuel implosion in inertial confinement fusion (ICF), for example. The RTI growth is also mitigated by the superimposition of the phase-controlled perturbations. [6-8, 15]

In ICF the fusion fuel compression is relevant to reducing an input driver energy.[2, 6, 16] In order to realize an ICF system, a uniform fuel implosion is essentially required to release a sufficient fusion output energy. For the uniform fuel implosion, the nonuniformity of the implosion acceleration should be less than a few %.[17, 18] The driver energy nonuniformity would induce a nonuniform energy deposition and implosion acceleration. The fuel implosion nonuniformity would degrade the uniform fuel compression and the fusion energy output. Ion beam accelerators can provide HIB axis oscillation with a high frequency of a few hundred MHz to ~GHz to realize the dynamic smoothing[5-8]. It has been found that the oscillating or wobbling motion of each HIB axis induces a phase-controlled HIBs energy deposition, and consequently the phase-controlled implosion acceleration is realized. Therefore, the HIBs illumination non-uniformity is successfully mitigated. Consequently, the RTI growth and the nonuniformity introduced by the driver HIBs are mitigated. The wobbling HIBs would provide an improvement in the fusion energy output gain.[6, 15, 19, 20] This result also demonstrates a viability of the dynamic mitigation method.

On the other hand, it has been found that the phase of the growing perturbations mitigated would be still defined by the initial imprint.[6-8, 14, 21] In this paper, the initial imprint effect is studied on the dynamic mitigation mechanism in plasmas. The results in this paper demonstrate that the initial imprint effect is successfully reduced by a pulse shaping of the oscillating or wobbling perturbation. In this paper we employ RTI as an example instability in plasmas.

In the next section, we summarize the dynamic smoothing mechanism in plasmas and shows the initial imprint effect on the dynamic mitigation. The results demonstrate that the dynamic smoothing mechanism would be further improved by reducing the initial imprint effect on the dynamic mitigation by the pulse shaping of the input driver.

## II. DYNAMIC MITIGATION MECHANISM

First, we summarize the dynamic mitigation mechanism in plasmas.[6-8] Instability grows from perturbations of physical quantities in unstable systems, and it would be difficult to measure the perturbation phase and amplitude in plasmas. Usually, the instability growth rate is employed to examine the plasma state. However, if we actively impose the perturbation phase by a driving energy source wobbling or oscillation, and so if we know or define the phase of the perturbations imposed actively, the perturbation growth would be controlled in a similar way as in the usual control mechanism[9].

For unstable plasmas, plasmas would move to nonlinear phase and then to turbulent state[22, 23]. When plasmas are in the turbulent stage, the phase of the turbulences may not be controlled and the dynamic mitigation theory presented here may not work for the plasma control. However, the dynamic mitigation method would work for pulse-operated plasmas, like ICF fuel target implosion, pulse-operated discharge plasmas, etc. In addition, if other sources of perturbations are introduced in plasmas and their phases are not controlled, the dynamic mitigation mechanism does not also work. The dynamic mitigation mechanism discussed here works, when the phase of the perturbation is defined or controlled actively from the outside.

Figure 1 shows an example unstable system, which has one mode of the wave number $k = 2\pi/\lambda$ with the amplitude of $a = a_0 e^{ikx+\gamma t}$ and with the growth rate $\gamma$ of the instability. Here $\lambda$ is the wave length of the perturbation. An example initial perturbation is shown in Fig. 1a). At *t*=0 the perturbation is imposed. The initial perturbation grows with $\gamma$. After $\Delta t$, if another perturbation which has an inverse phase is actively imposed as shown in Fig. 1b), the overall amplitude is the superimposition of all the perturbations, and the actual perturbation amplitude is well mitigated as shown in Fig. 1c).

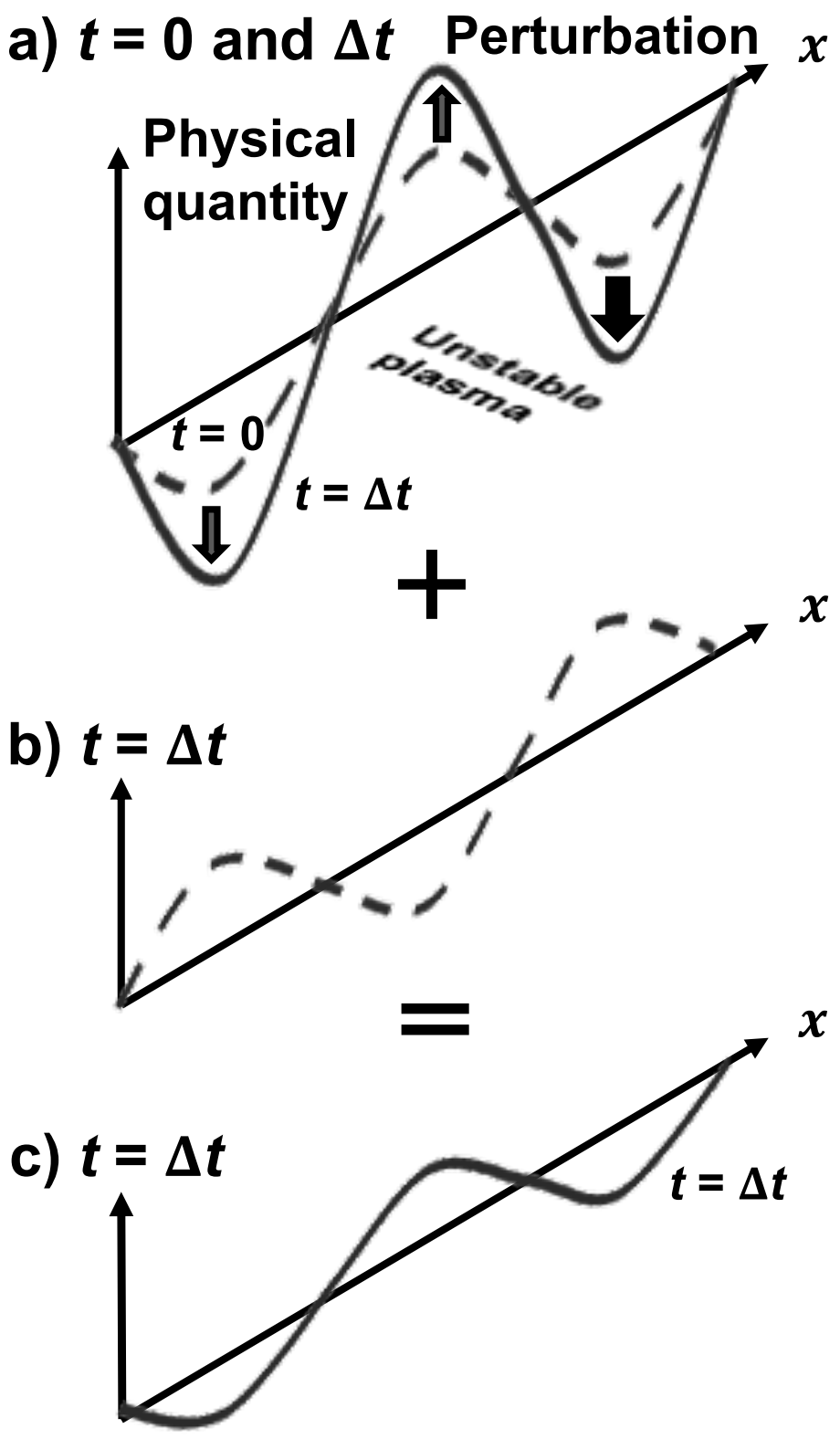

Fig. 1 Dynamic mitigation mechanism. a) At *t*=0, a perturbation with the wavelength of $\lambda$ is imposed. The perturbation grows with the growth rate of $\gamma$. b) After $\Delta t$, another perturbation, which has an inverse phase, is actively imposed. c) After the superimposition of the perturbations in a) and b) at $\Delta t$, the perturbation amplitude is mitigated well.

When the instability driver oscillates uniformly in time, the imposed perturbation for a physical quantity *F* at *t*=$\tau$ may be written as

$$F = \delta F e^{i\Omega\tau} e^{\gamma(t-\tau)+i\vec{k}\cdot\vec{x}}. \quad (1)$$

Here $\delta F$ is the amplitude, $\Omega$ the wobbling or oscillation frequency defined actively by the driving beam, and $\Omega\tau$ the phase shift of superimposed perturbations. At each time $t$, the wobbler or the modulated driver provides a new perturbation with the phase and the amplitude actively defined by the driving beam itself. The superimposition of the perturbations provides the overall perturbation at $t$ as follows:

$$\int_0^t d\tau\ \delta F e^{i\Omega\tau} e^{\gamma(t-\tau)+i\vec{k}\cdot\vec{x}} \propto \frac{\gamma+i\Omega}{\gamma^2+\Omega^2} \delta F e^{\gamma t} e^{i\vec{k}\cdot\vec{x}} \quad (2)$$

When $\gamma \leq 0$, the system is stable and Eq. (2) shows a simple dynamic smoothing of the perturbations. When $\gamma \geq 0$ and $\Omega \gg \gamma$ for an unstable system, the perturbation amplitude is reduced by the factor of $\sim \gamma/\Omega$, compared with that in the pure instability growth ($\Omega = 0$) based on the energy deposition nonuniformity[6-8]. When $\Omega \cong \gamma$ , the amplitude mitigation factor is still significant. The result in Eq. (2) presents that the perturbation phase should oscillate with $\Omega \gtrsim \gamma$ for the effective amplitude reduction.

Figures 2 show an example simulation results for RTI. In this case two modes of the initial perturbations are imposed with the normalized wavelength of $\lambda = 1$ and 1/2 in $x$ and $z$. Two uniform fluids are stratified under an acceleration of $g = g_0 + \delta g$. The normalized densities are 1 and 2 for the two plasmas, which are sustained by the acceleration $g_0$ imposed in the $-x$ direction. In this example, $\delta g$ introduces the perturbation. The amplitude of $\delta g$ is 5%, and the amplitude ratio between the two modes of $\lambda = 1$ and 1/2 is 1 : 1. Figures 2 show a) the density profile and b) the velocity at $\gamma t = 10.8$. Here the time $t$ is normalized by $1/\gamma$, where $\gamma$ is the RTI growth rate of $\gamma = \sqrt{gk}$ and the normalized wavenumber $k = 2\pi/(1/2) = 4\pi$. In this paper the dynamic oscillation or wobbling is introduced as $e^{i\Omega\tau}$, where $\Omega$ shows the wobbling frequency. When $\Omega = 0$ as shown in Figs. 2, the RTI grows as usual. The displacement $H$ of the interface between two stratified fluids is measured in this paper.

In Fig. 2a) the displacement $H$ is indicated for the RTI growth without the dynamic mitigation ($\Omega = 0$). In Fig. 3 the oscillation is introduced in $\delta g$ with $\Omega = \gamma$. Figure 3 shows the density profile at $\gamma t = 10.8$, and the mitigation in $H$ is realized. Figure 4 shows the histories of $H$ for both the cases. The reduction ratio is measured by the ratio of $(H_{\Omega=0} - H)/H_{\Omega=0}$ and is 42.0% in $H$ between the two cases in Figs. 2 and 3.

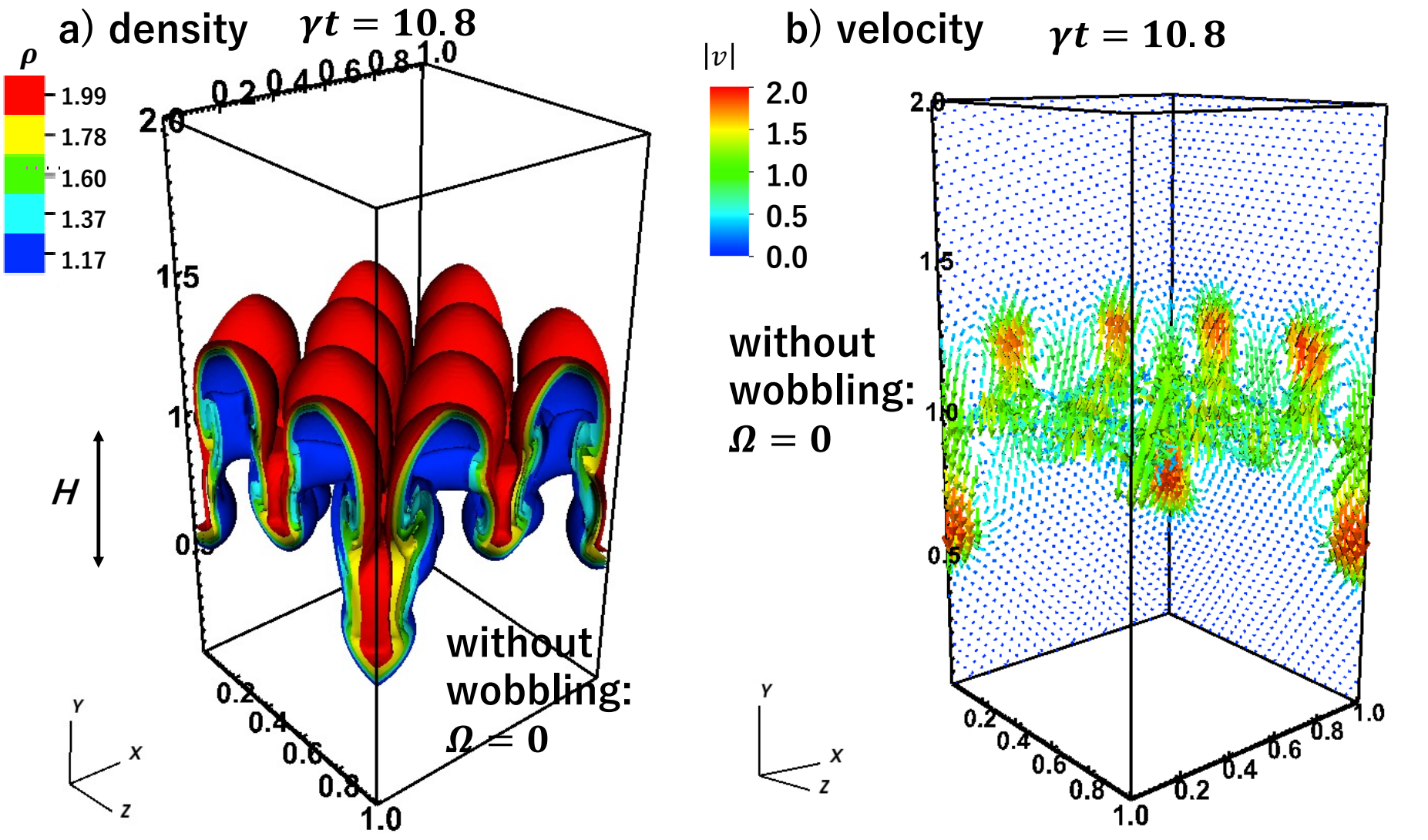

Fig. 2. Rayleigh-Taylor instability (RTI) growth under an acceleration perturbation of $\delta g$: a) The density profile and b) the velocity at $\gamma t = 10.8$. Two modes of the initial perturbations are imposed with the normalized wavelength of $\lambda = 1$ and 1/2 in $x$ and $z$. Two uniform fluids are stratified under an acceleration of $g = g_0 + \delta g$. The normalized densities are 1 and 2 for the two plasmas, which are sustained by the acceleration $g_0$ imposed in the $-x$ direction. The amplitude of $\delta g$ is 5%, and the amplitude ratio between the two modes of $\lambda = 1$ and 1/2 is 1 : 1. Here the time $t$ is normalized by $1/\gamma$, where $\gamma$ is the RTI growth rate of $\gamma = \sqrt{gk}$ and the normalized wavenumber $k = 2\pi/(1/2) = 4\pi$.

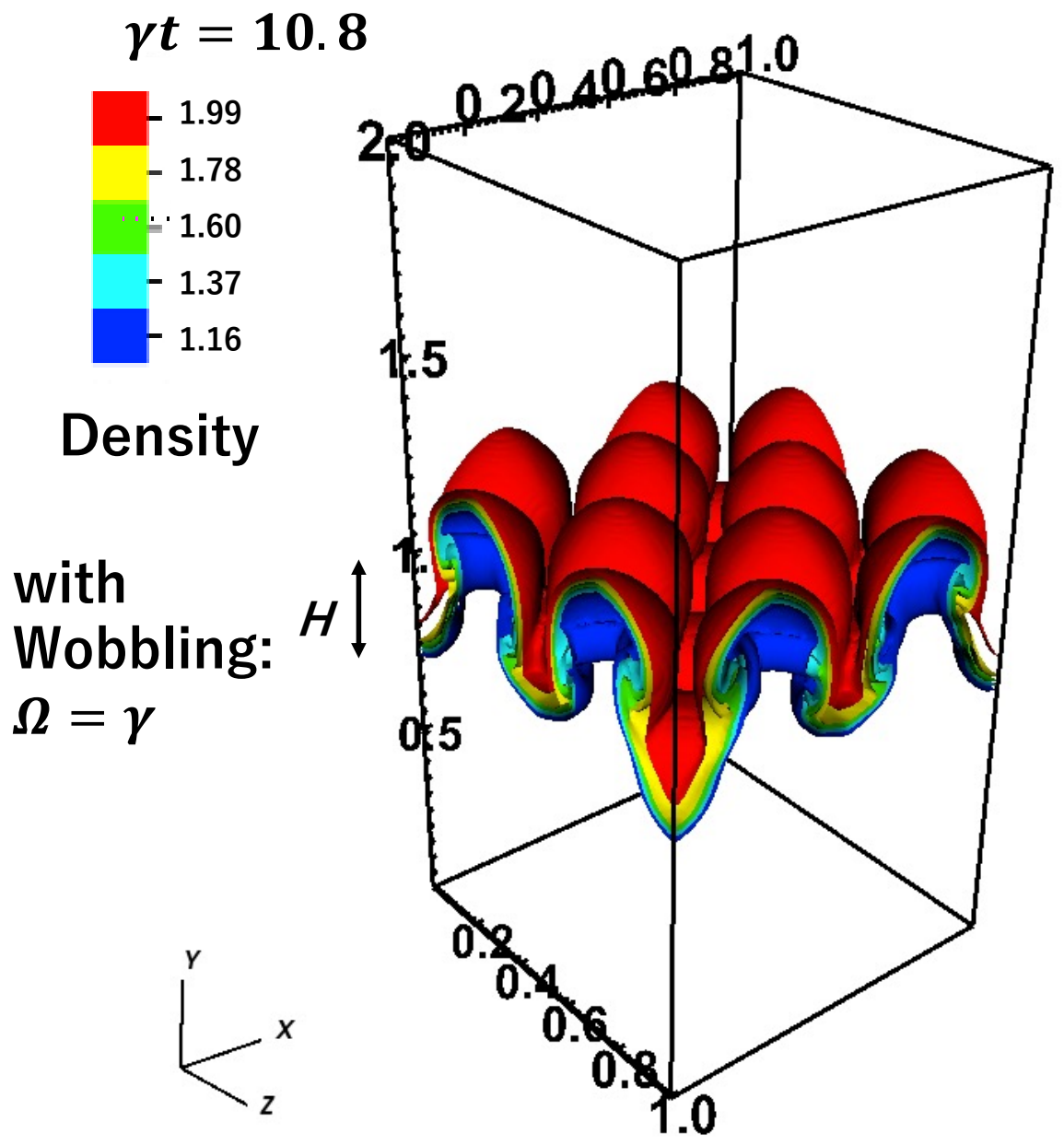

Fig. 3. RTI growth under the dynamic mitigation. The density profile is shown at $\gamma t = 10.8$ under the oscillation in $\delta g$ with $\Omega = \gamma$, and the dynamic mitigation in $H$ is realized.

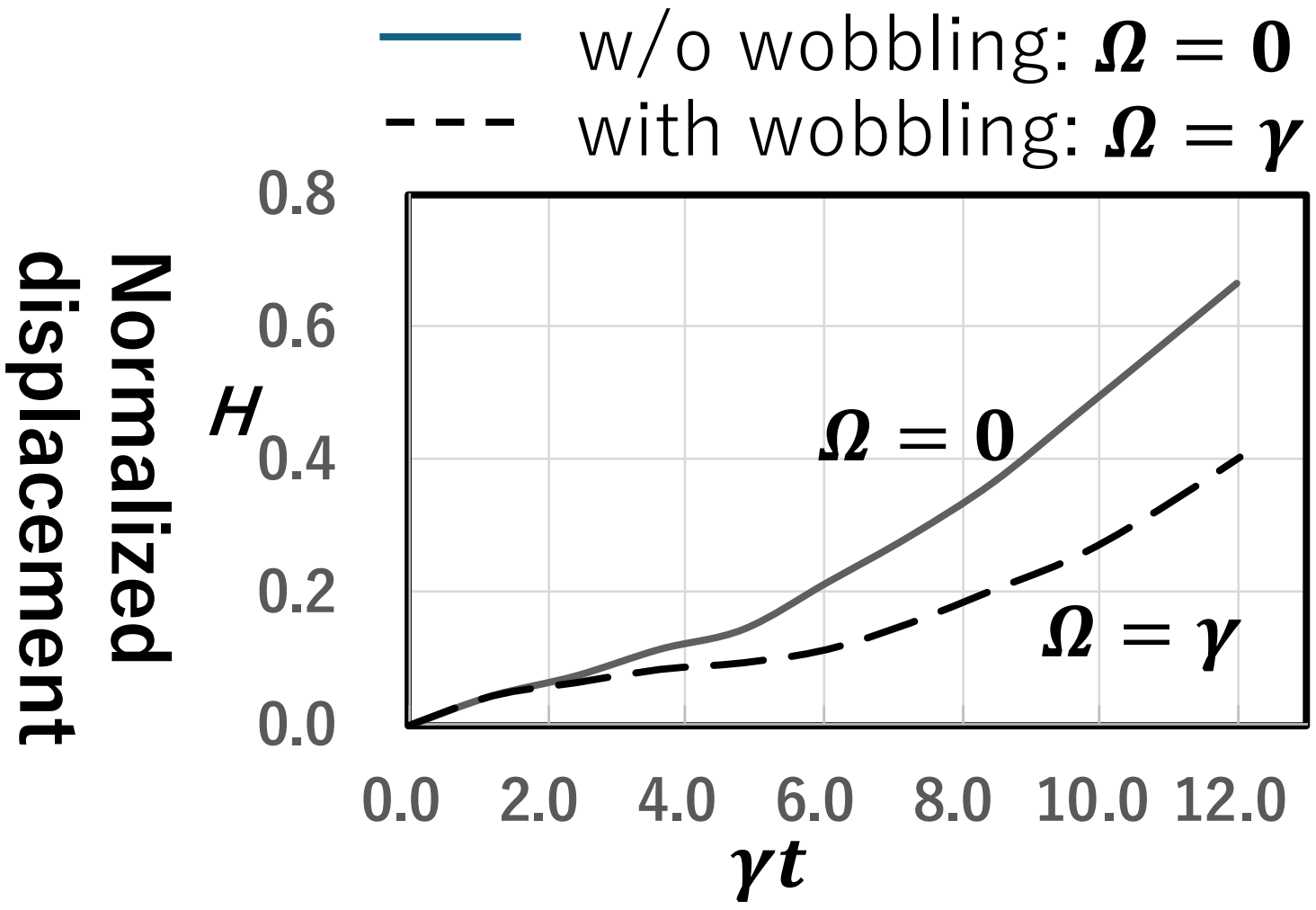

Fig. 4. Histories of $H$ for the two cases in Figs. 2 ($\Omega = 0$) and 3 ($\Omega = \gamma$). The reduction ratio is measured by the ratio of $(H_{\Omega=0} - H)/H_{\Omega=0}$ and is 42% in $H$ between the two cases. In the early time the perturbation in the case of $\Omega = \gamma$ follows the growth curve of the perturbation with $\Omega = 0$. This would suggest that the initial imprint still defines the perturbation mode and phase even in the case of $\Omega = \gamma$.

Figures 2-4 demonstrate that the dynamic mitigation mechanism is viable to reduce the instability growth effectively. At this point, if we focus on the phase of the perturbation in Fig. 3, one can find that the phase of the mitigated perturbation is quite similar to the phase in Figs. 2. This means that the perturbation phase would be still defined by the initial perturbation phase. In Fig. 4, in the early time the perturbation in the case of $\Omega = \gamma$ follows the growth curve of the perturbation with $\Omega = 0$. Figure 5 shows the mode analyses for the two cases. At $\gamma t = 10.8$, the mode 2 becomes dominant with the mode 1 and the mode 3 for both the two cases. The results shown in this Section suggest that the initial imprint still defines the perturbation mode and phase. The results may also suggest us to find a better way to employ the dynamic mitigation mechanism by reducing the initial imprint effect on the mechanism. This point is studied in the next Section.

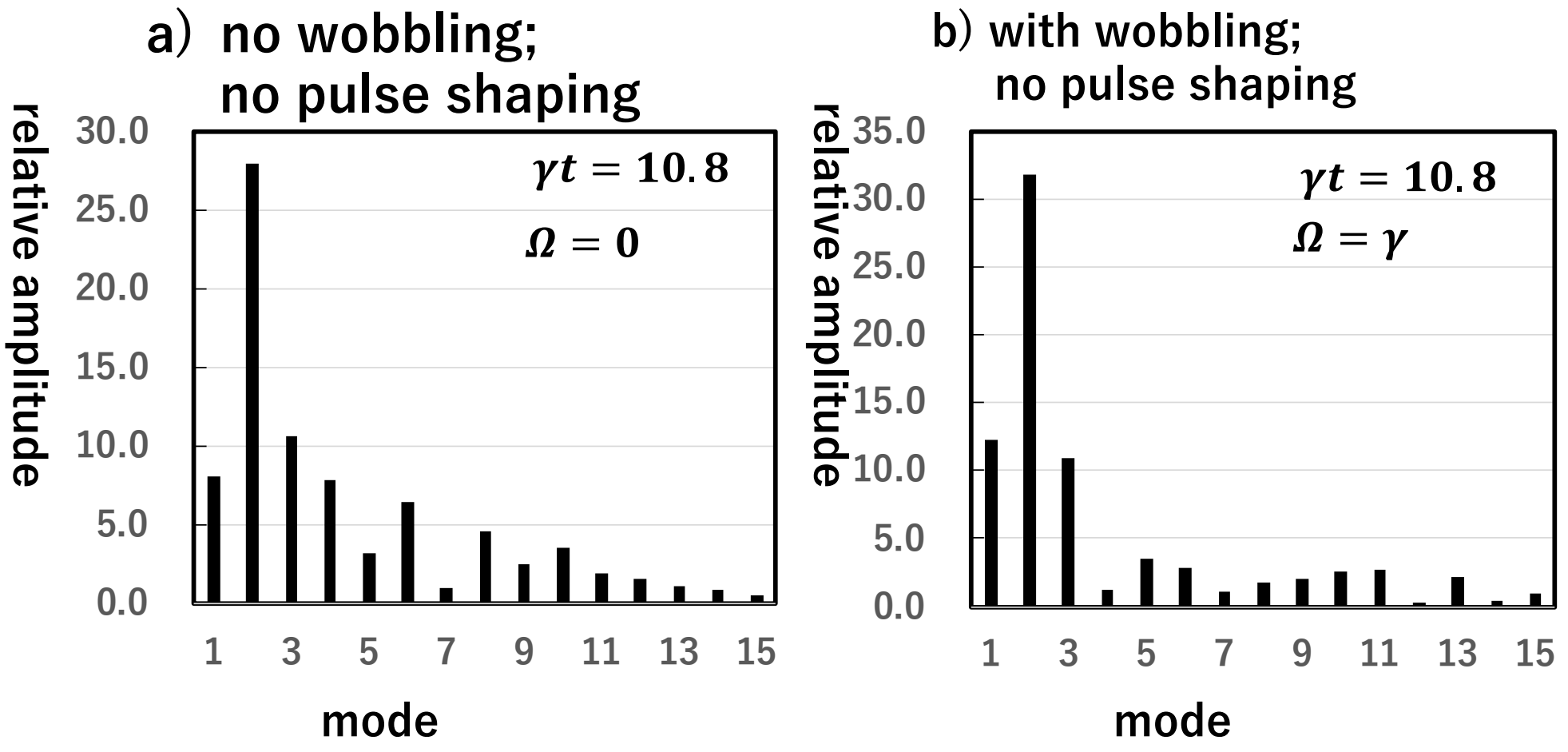

Fig. 5. Mode analyses for the two cases in Figs. 2 ($\Omega = 0$) and 3 ($\Omega = \gamma$) at $\gamma t = 10.8$. The perturbation mode and phase would be still defined by the initial perturbation mode and phase.

## III. REDUCTION OF INITIAL IMPRINT EFFECT ON DYNAMIC MITIGATION

In the last Section we found that the initial imprint effect remains in the perturbation mode and phase even under the dynamic mitigation. The initial imprint effect could be also found in filamentation instability, the Weibel instability, kink and sausage instabilities of magnetized column, two-stream instability, tearing mode instability and the Kelvin-Helmholtz instability, as well as RTI instability [6-8, 12, 14-16]. It would be easily found that if the oscillation or wobbling frequency $\Omega$ is much higher compared with the instability growth rate $\gamma$, the initial imprint effect is reduced. However, it would not be always possible to realize $\Omega \gg \gamma$, depending on the dynamic oscillation or wobbling capability.

If we look back at Eq. (1), the amplitude of the perturbation $\delta g$ was set to be constant in the two cases in Fig. 2 and 3. As we discussed before on the situation in which the dynamic mitigation mechanism works well, the dynamic mitigation mechanism would be effective for pulsed-operated plasmas. In pulsed plasmas the pulse rise time of $\tau_r$ would be finite as shown in Fig. 6. In the last Section the amplitude of $\delta g$ was constant in time, that means, a square pulse with $\tau_r = 0$. In this Section we employ the finite $\tau_r$ for the amplitude of the perturbation $\delta g(t)$. In this Section a sigmoid function of $1/\left(1 + e^{-3(3\gamma t-2)}\right)$ is used to describe the pulse shape. In this pulse shape the pulse rise time is $\tau_r \sim 1/\gamma$. In order to reduce the initial imprint effect on the dynamic mitigation mechanism, the conditions of $\tau_r \Omega \geq 1$ and $\tau_r \gamma \geq 1$ should be satisfied. In ICF, especially in HIB ICF (HIF), these conditions are satisfied [6, 15, 16, 19].

Figures 7 present the RTI growths for the cases of a) $\Omega = 0$ and b) $\Omega = \gamma$ with the pulse shaping shown above with $\tau_r \sim 1/\gamma$. Figure 7b) shows that the perturbation modes are modified and the amplitude is reduced successfully by the wobbling motion. Figure 8 demonstrates that the perturbation growth is well mitigated by the dynamic mitigation mechanism. It is clearly shown that in the early stage the perturbations in the cases of $\Omega = \gamma$ and $\Omega = 10\gamma$ do not follow the perturbation growth in the case of $\Omega = 0$. The mitigation ratios, measured by $(H_{\Omega=0} - H)/H_{\Omega=0}$, are 61.0% and 80.0% in $H$ at $\gamma t = 10.9$ for $\Omega = \gamma$ and $\Omega = 10\gamma$, respectively. Figure 9 presents the mode analyses for the cases of a) $\Omega = 0$ and b) $\Omega = \gamma$. The perturbation modes are significantly modified from those in Fig. 9a). The results shown here would present that the pulse shaping contributes to reduce the initial imprint effect on the dynamic mitigation mechanism.

When the rise time of $\tau_r$ becomes short to $\tau_r \sim 1/(2\gamma)$ and other parameter values are same with those in Fig. 7b), the condition of $\tau_r \gamma \geq 1$ is violated, and for $\Omega = \gamma$ the mitigation ratio becomes small compared with that in Fig. 7b) by 23.4% at $\gamma t = 10.9$. The perturbation mode is also similar to that in Fig. 9a), and the initial imprint effect is recovered in this case as expected.

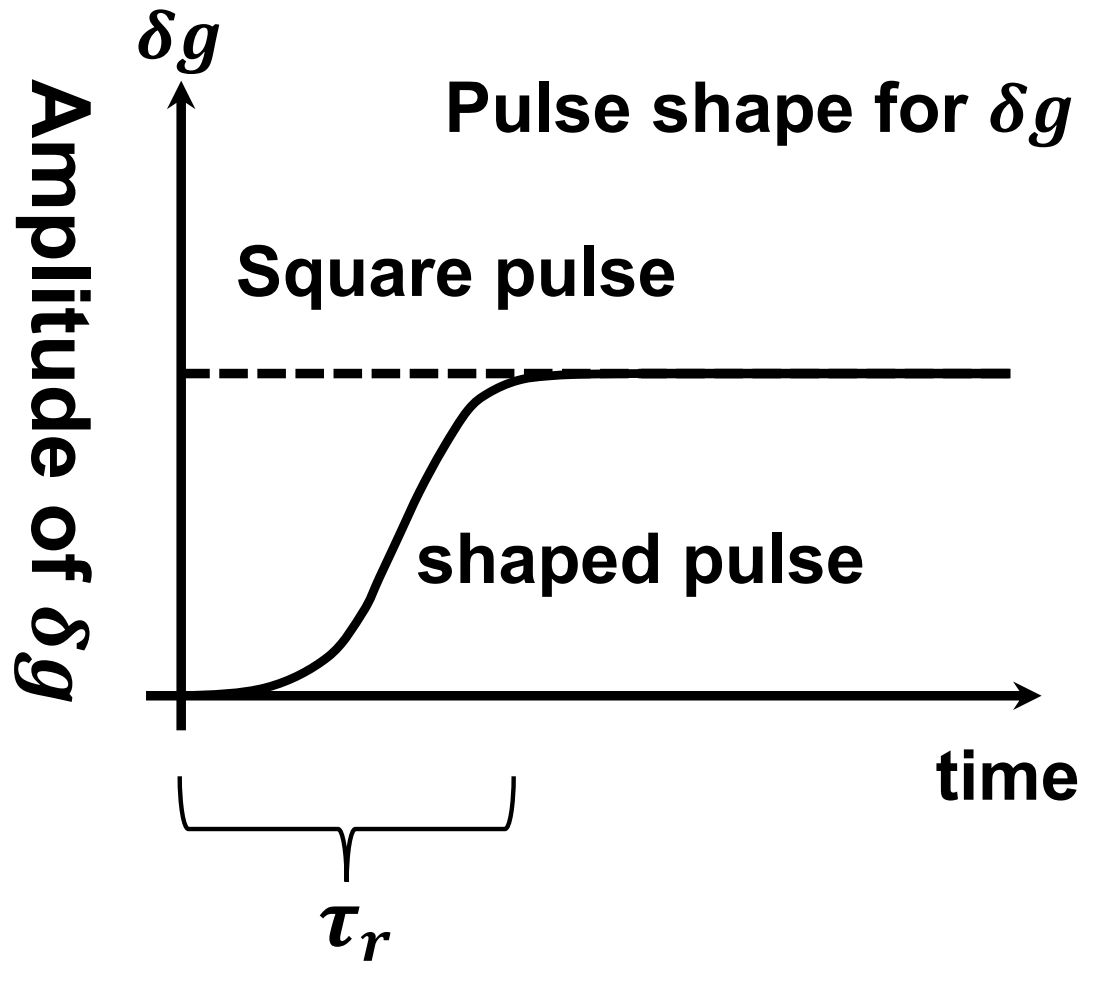

Fig. 6. Finite rise time $\tau_r$ of shaped pule would reduce the initial imprint effect on the dynamic mitigation mechanism.

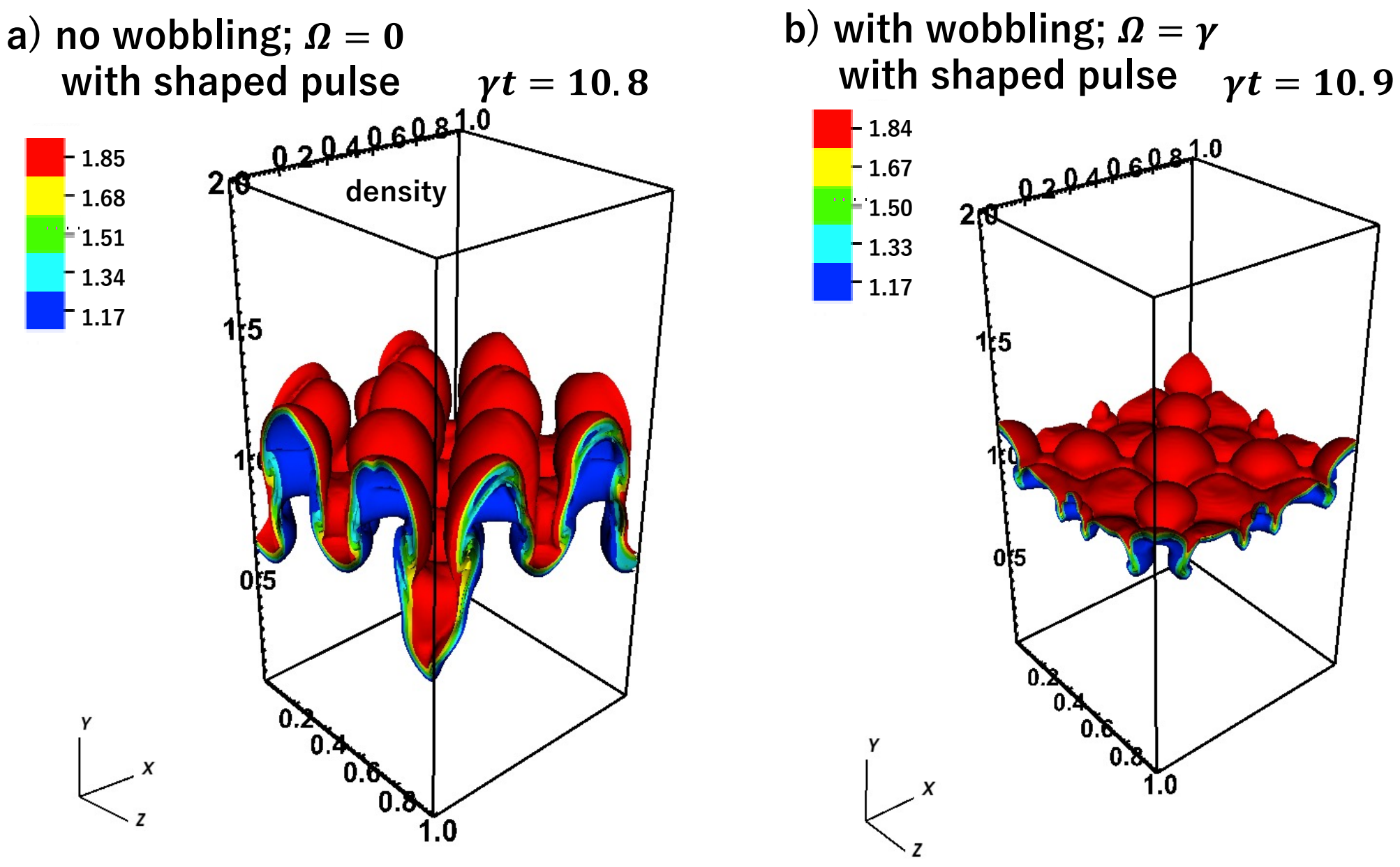

Fig. 7. RTI growths for the cases of a) $\Omega = 0$ and b) $\Omega = \gamma$ with the pulse shaping shown in Fig. 6. Figure 7b) shows that the perturbation mode is modified, and the amplitude is reduced successfully by the wobbling motion.

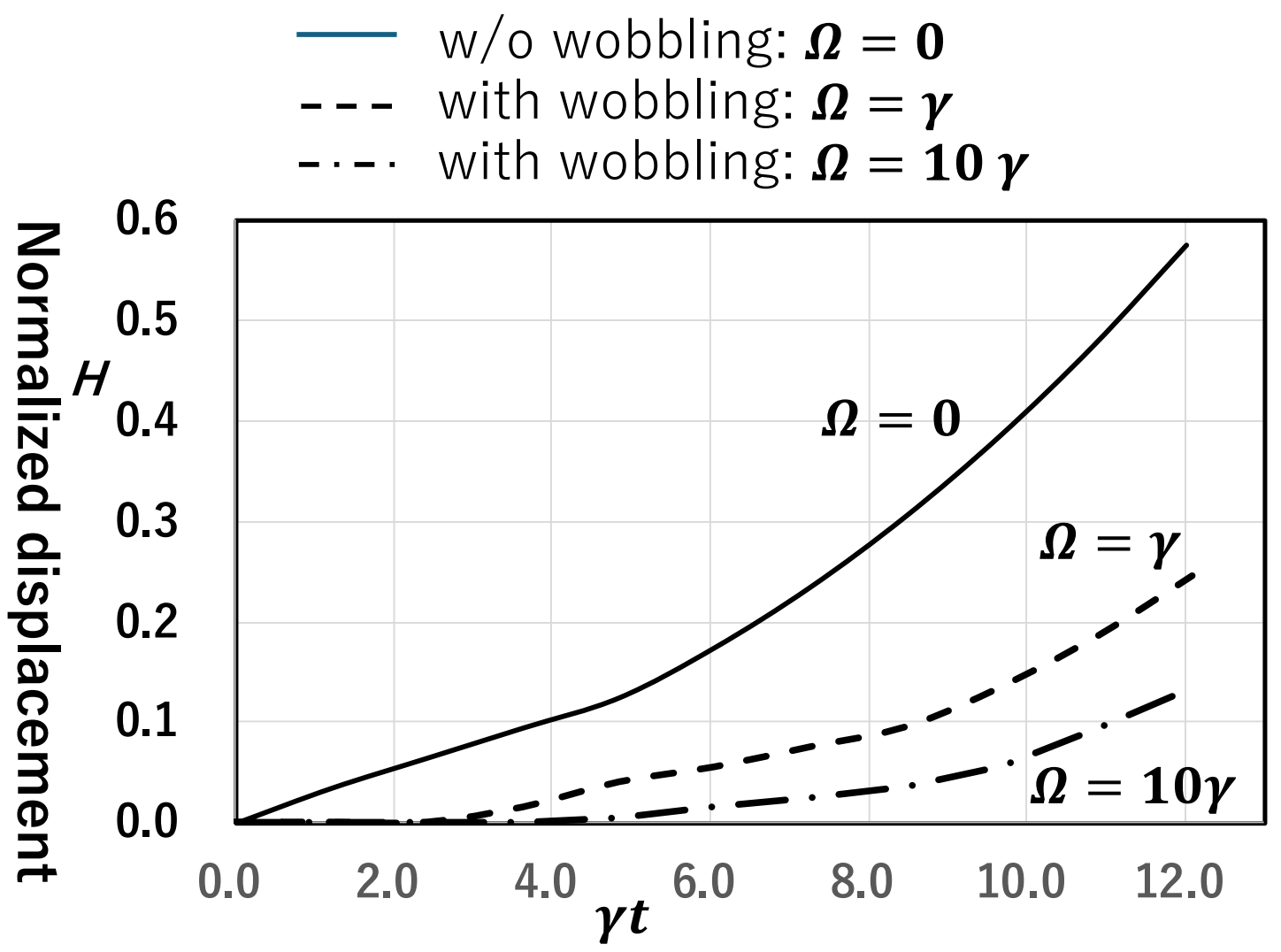

Fig. 8. Histories of the perturbation displacement $H$. The perturbation growth is well mitigated by the dynamic mitigation mechanism. It is clearly shown that in the early stage the perturbations in the cases of $\Omega = \gamma$ and $\Omega = 10\gamma$ do not follow the perturbation growth in the case of $\Omega = 0$. The reduction ratios, measured by $(H_{\Omega=0} - H)/H_{\Omega=0}$, are 61.0% at $\gamma t = 10.8$ and 80.0% at $\gamma t = 10.9$ in $H$ for $\Omega = \gamma$ and $\Omega = 10\gamma$, respectively.

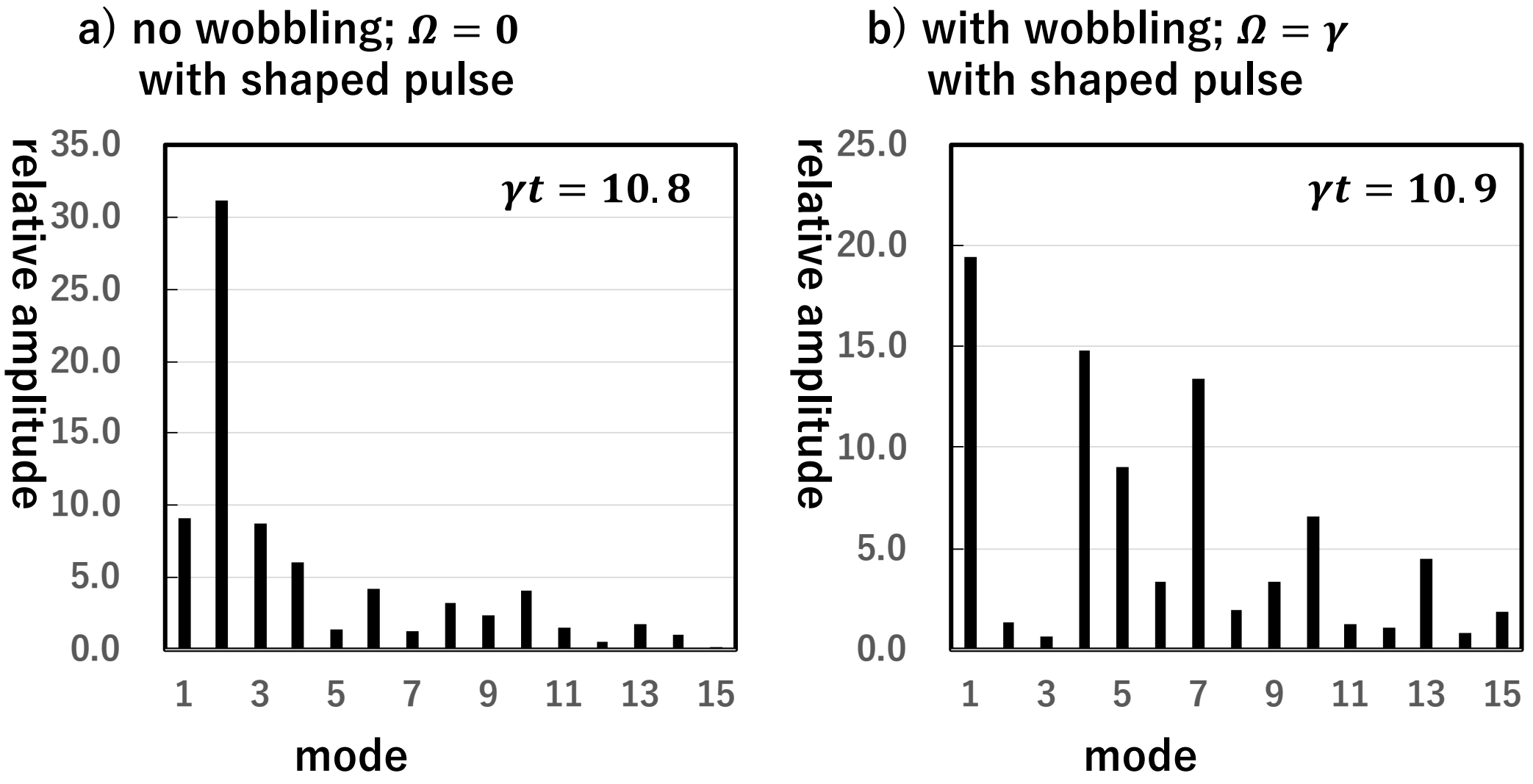

Fig. 9. Mode analyses for the two cases shown in Figs. 7a) ($\Omega = 0$) at $\gamma t = 10.8$ and b) ($\Omega = \gamma$) at $\gamma t = 10.9$. The perturbation modes are significantly modified from those in Fig. 9a) ($\Omega = 0$). The pulse shaping contributes to reduce the initial imprint effect on the dynamic mitigation mechanism.

## IV. CONCLUSIONS

We have presented the initial imprint effect on the dynamic mitigation mechanism and found its reduction method in plasma instability. The initial imprint effect remains in the dynamic mitigation mechanism under a square pulse with $\tau_r = 0$ in Fig. 6, and in this case the initial imprint still defines the perturbation phase and mode in plasmas, even when $\Omega \geq \gamma$. However, when the driver perturbation pulse is shaped with a finite rise time of $\tau_r$, the perturbation growth is reduced well and the dynamic mitigation mechanism successfully reduces the initial imprint effect on the instability growth. The driver pulse shaping would be introduced intentionally or unintentionally, for example, in pulses in ICF fuel target driver [1, 6], plasma heating, etc.

The dynamic smoothing and stabilization mechanism employed in this paper is based on the phase control of the non-uniformities introduced. Even in plasmas we can control the phases of the perturbations which are actively imposed from the outside of the systems. For instance, the drivers, like the wobbling HIBs in HIF, would introduce their own non-uniformities. The phases of the non-uniformities imposed by the drivers are controlled, for example, by the wobbling motion, and consequently the dynamic stabilization or smoothing mechanism would be implemented in plasmas. The dynamic stabilization and smoothing mechanism would be applied to various plasma instabilities and non-uniformities, as far as the phase of the non-uniformities are controllable.

We would like to point out again that the dynamic stabilization and smoothing mechanism is not almighty. If the non-uniformity phase cannot be controlled actively, the mechanism does not work in plasma instabilities. For example, if an ICF fuel pellet has an unexpected aspherical shape or if the fusion pellet has a non-uniformity in the shell thickness, the dynamic stabilization and smoothing mechanism does not work.

**ACKNOWLEDGEMENTS**

The work was partly supported by JSPS, MEXT and Laboratory for Laser Plasmas, Shanghai Jiao Tong University.

**Author contributions statements**

SK proposed the basic idea for the control method and its application in this paper. SK prepared the manuscript text and the figures. SK prepared the computer simulation software used in this paper.

**Additional Information:**

**Competing interests:** The authors declare no competing interests.

**Availability of Data:** The data that supports the findings of this study are available within the article

**Software:** The computer software used in this paper is available in Ref. 20 and partly in Ref. 6.